\def\eeq{\end{equation}}
\def\beq{\begin{equation}}
\def\bea{\begin{eqnarray}}
\def\eea{\end{eqnarray}}

\documentstyle[aps,prl,epsfig]{revtex}

\begin{document}

\title{Exact and approximate results of non-extensive quantum statistics}
\author{U\v{g}ur T{\i}rnakl{\i}$^{1,2,}$\thanks{e-mail:
tirnakli@sci.ege.edu.tr}
and
Diego F. Torres$^{3,}$\thanks{e-mail: dtorres@venus.fisica.unlp.edu.ar}}

\address{$^1$ Centro Brasileiro de Pesquisas Fisicas, Rua Dr. Xavier
Sigaud 150, 22290-180 Rio de Janeiro, Brazil \\
$^2$ Department of Physics, Faculty of Science, Ege University
35100 Izmir-Turkey\\
$^3$ Departamento de F\'{\i}sica,
Universidad Nacional de La Plata,
C.C. 67, 1900, La Plata,  Buenos Aires, Argentina}

\maketitle

\begin{abstract}

We develop an analytical technique to derive explicit forms of
thermodynamical quantities within the asymptotic approach to
non-extensive quantum distribution functions. Using it, we find an
expression for the number of particles in a boson system which we
compare with other approximate scheme (i.e. factorization
approach), and with the recently obtained exact result. To do
this, we investigate the predictions on Bose-Einstein condensation
and the blackbody radiation. We find that both approximation
techniques give results similar to (up to ${\cal O}(q-1)$) the
exact ones, making them a useful tool for computations. Because of
the simplicity of the factorization approach formulae, it appears
that this is the easiest way to handle with physical systems
which might exhibit slight deviations from extensivity.

\noindent {\it PACS Number(s):} 05.20.-y, 05.30. Jp, 05.30.Fk

\end{abstract}


\section{Introduction}
Since the papers by Tsallis \cite{first,second}, non-extensive
statistical formalism has been shown to be not only {\it robust}
--it allows generalizations of all necessary fundamental concepts
of thermostatistics \cite{concepts}--, but also {\it useful} --it
provides a suitable theoretical tool to explain some of the
experimental situations where standard thermostatistics has
shortcomings, due to the presence of long-range interactions, or
long-range memory effects, or (multi)-fractal space-time
constraints--. See Ref. \cite{biblio} for a periodically updated
bibliography. \\

The core of this generalized formalism is defined through a
generalized entropy \beq S_q = k \frac{1- \sum_{i=1}^{W}
p_i^q}{q-1}, \;\;\;\;\;\; (q \in \Re), \eeq where $k$ is a
positive constant, $\{p_i\}$ is a set of probabilities and $W$ is
the total number of microscopic configurations. It is easy to
verify that the $q\rightarrow 1$ limit immediately recovers the
usual (extensive) Boltzmann-Gibbs entropy. Moreover, if a composed
system $A+B$ has probabilities which factorize into those
corresponding to the subsystems $A$ and $B$, then $S_q(A+B)/k =
S_q(A)/k + S_q(B)/k + (1-q) S_q(A)S_q(B)/k^2$. This property
clearly exhibits the fact that the parameter $q$ characterizes the
degree of non-extensivity of any physical system.\\

The generalization of quantum statistics for non-extensive systems
was only accomplished, up to recent days, in an {\it approximate
fashion}, by using two different schemes. One of them is the
Asymptotic Approach (AA), of Tsallis et al. \cite{bbAAtsallis},
the other one is the Factorization Approach (FA), of
B\"{u}y\"{u}kk{\i}l{\i}\c{c} et al. \cite{FA}. The physical
applications studied so far within these two approximations
include the blackbody radiation
\cite{bbAAtsallis,bbAAroditi,bbFAugur}, the Stefan-Boltzmann
constant \cite{SB-AAplastino,SB-AAwang,SB-FAugur}, and some
aspects of the early universe physics \cite{early-AA,early-FA}.
Moreover, the AA has also been used in some other works such as
the Bose-Einstein condensation \cite{curilef}, the specific heat
of $^4$He \cite{curilef-helyum}, thermalization of an
electron-phonon system \cite{koponen} and cosmology
\cite{torres1,torres2}. Although some detailed analysis on these
approximate schemes \cite{wang-FA} suggest that both schemes could
be helpful in physical applications --at least, for $(1-q)$-order
corrections--, this was still doubtful. A complete verification
needed a comparison between the results of these approximate
schemes and the exact ones. But an exact treatment of
non-extensive quantum distributions was not available up to the
recent papers of Rajagopal et al. \cite{raj,ervin}. Just after
this work, Lenzi and Mendes have also given an exact treatment of
blackbody radiation \cite{bb-exact}. All these recent efforts
enable us to make a comparison between the approximate and exact
schemes, which will ultimately show whether the AA and the FA are
useful or not. This will be the main purpose of this paper.\\

In Section II, we review the approximate and exact results and
develop an analytical method to derive the explicit form of any
measurable quantity  within the AA. We compare the approximate and
exact results in Section III using (i) the predictions of one of
the experimental tests suggested in \cite{raj} and (ii) the
blackbody radiation. Finally, we give our final comments in
Section IV.

\section{Non-extensive quantum statistics}

\subsection{Asymptotic approach}

Within the AA, namely in the $\beta(1-q)\rightarrow 1$ limit, the
generalized partition function is given by \cite{bbAAtsallis}

\beq Z_q\simeq Z_{1} \left\{1-\frac 12 (1-q) \beta^2 \left<\hat
{\cal H}^2\right>_{1}\right\}, \eeq from where the generalized
distribution function of non-interacting bosons can be found, up
to $(1-q)$-order, as \beq \label{nAA} \left<n\right>_q
=\left<n\right>_1 + (1-q) \left<n\right>_1 \left\{\ln(1/Z_1) +
(x-\psi) \left[\frac{\left<n^2\right>_1}
{\left<n\right>_1}+(x-\psi)\left(\left<n^2\right>_1 -
\frac{\left<n^3\right>_1}{\left<n\right>_1}\right)\right]\right\},
\eeq where $x\equiv\beta\epsilon$ , $\psi\equiv\beta\mu$, and \beq
\left<n\right>_1 = \frac{1}{e^{x-\psi}-1},\hspace{0.4cm}
\left<n^2\right>_1 = \frac{e^{-(x-\psi)}+e^{-2(x-\psi)}}
{\left[1-e^{-(x-\psi)}\right]^2},\hspace{0.4cm} \left<n^3\right>_1
= \frac{e^{-(x-\psi)}+ 4e^{-2(x-\psi)}+e^{-3(x-\psi)}}
{\left[1-e^{-(x-\psi)}\right]^3}. \eeq The standard ($q=1$)
partition function is given by \beq Z_1 =
\frac{1}{1-e^{-(x-\psi)}}. \eeq This approximation has found a
wide range of applications up to now, however, no attempt has been
made for deriving some of the thermodynamical quantities within
this approach, directly using Eq. (\ref{nAA}).
\\

One aim of this paper is to provide a technique for computing, in
a closed form, the kind of integrals needed to find the average
number of particles within the AA. To do this, let us start by
writing down the definition of the average number of particles:
\beq \label{bosN} \left<N\right>_q = \frac{2\pi V(2mk)^{3/2}
T^{3/2}}{h^3} \int_0^{\infty} \epsilon^{1/2} \left<n\right>_q
d\epsilon , \eeq where all variables have the usual meaning. Using
Eq. (\ref{nAA}) and the definitions of $x$ and $\psi$, this
expression turns out to be \beq \left<N\right>_q = \frac{2\pi
V(2mk)^{3/2} T^{3/2}}{h^3} \left[I_{st} + (1-q)(I_2 + I_3)
\right], \eeq where \beq I_{st} = \int_0^{\infty}\frac{x^{1/2}
dx}{e^{x-\psi}-1},\hspace{0.4cm} I_{2} =
\int_0^{\infty}\frac{(x-\psi) x^{1/2} dx} {e^{x-\psi}-1}, \eeq and
\beq I_{3} = \int_0^{\infty}\frac{(x-\psi) x^{1/2} dx}
{e^{x-\psi}-1} \left[\frac{\left<n^2\right>_1}
{\left<n\right>_1}+(x-\psi)\left(\left<n^2\right>_1 -
\frac{\left<n^3\right>_1}{\left<n\right>_1}\right)\right]. \eeq
$I_2$ and $I_3$ are the $(q-1)$ order correction to the standard
($q=1$) result and here $I_{st}$ stands for the standard integral
appearing in the solution of the extensive case \cite{pathria}.
$I_{st}$ and $I_2$ have standard forms,  and could easily be
solved as: \beq I_{st} = \Gamma(3/2) g_{3/2}(z), \eeq \beq
\label{2RESUL} I_2 = \int_0^{\infty} \frac{x^{5/2-1}
dx}{e^{x-\psi}-1} - \psi \int_0^{\infty} \frac{x^{3/2-1}
dx}{e^{x-\psi}-1} = \Gamma(5/2) g_{3/2}(z) - \psi \Gamma(3/2)
g_{3/2}(z) , \eeq where $z$ is the fugacity and is defined as
$z\equiv e^{\beta\mu}$. On the other hand, $I_3$ is more involved,
and it takes the form: \beq \label{3RESUL} I_3 = a+b-3c-d, \eeq
where \beq a= \int_0^{\infty} \frac{(x-\psi)x^{1/2}  dx}
{\left[e^{x-\psi}-1\right]^2}, \hspace{0.4cm} b= \int_0^{\infty}
\frac{(x-\psi)x^{1/2} e^{x-\psi} dx}
{\left[e^{x-\psi}-1\right]^2}, \eeq \beq c=\int_0^{\infty}
\frac{(x-\psi)^2 x^{1/2} e^{x-\psi} dx}
{\left[e^{x-\psi}-1\right]^3}, \hspace{0.4cm} d=\int_0^{\infty}
\frac{(x-\psi)^2 x^{1/2} e^{2(x-\psi)} dx}
{\left[e^{x-\psi}-1\right]^3}. \eeq In an Appendix, we provide an
analytical technique (maybe there are others) to compute each one
of these integrals. Using this technique, we obtain the average
number of particles as: \bea \left<N_e\right>_q \frac {h^3}{2\pi
V(2mkT)^{3/2}} = \Gamma(3/2) g_{3/2}(z) + (q-1) \sqrt{\pi} \times
\hspace{7cm} \mbox{}    \nonumber\\ \hspace{3cm} \times \left[
\frac 32 g_{3/2}(z)  -\frac 98 g_{5/2}(z) + \frac {7}{4} \psi
g_{3/2}(z) - 2 \psi g_{1/2}(z) - \frac 12 \psi^2 g_{1/2}(z) +
\frac 98 \psi^2 g_{-1/2}(z) \right]. \eea Here,
$\left<N_e\right>_q$ stands for the number of particles in the
excited states ($\epsilon\neq 0$). As in the standard case, we
have separated the contribution of the state given by $\epsilon
=0$, which has zero weight in the integrals. For this level of
energy, we found, \beq \left<N\right>_q (\epsilon = 0) =
\frac{z}{1-z} \left\{ 1 + (q-1) \left[ \ln z +
 \frac{z \ln z}{1-z} - \frac{ \ln z}{1-z}
 + 3 \frac{z (\ln z)^2}{(1-z)^2}
+ \frac{(\ln z)^2}{(1-z)^2}\right] \right\} . \eeq When $z\ll 1$,
all the correction terms go to zero. When $z\rightarrow 1$, some
of the terms are divergent but the usual shape is unchanged. This
can be seen in Fig. 1.\\

Numerical analysis, which we show in Fig. 2, illustrates that the
maximum correction is attained for $z=1$. Then, the number of
particles in all excited states is bounded by, \beq \label{neAA}
\left<N_e\right>_q \le \frac {2\pi V(2mk)^{3/2} T^{3/2}} {h^3}
\left[2.315 + (q-1) 4.27 \right] . \eeq It is worth noticing that
the AA is such that not all terms in the $(1-q)$ correction are
positive (or negative, depending on the choice of $q$) definite.
Moreover, their maximum values are not always attained at middle
points of the interval of interest, and although they do have
bounded expressions, the maximum correction is obtained only for
$z=1$. This differs from what happened in the FA, where each term
had a maximum value within the interval of interest
\cite{tor-tir}. The order of magnitude of the maximum correction
is, however, the same in both approximations.\\

Any interested reader could easily apply the same technique, which
we introduced in the Appendix, to compute any other
thermodynamical quantity, whenever it is needed.

\subsection{Factorization approach}

Within the FA \cite{FA}, the generalized distribution function of
bosons is given, at $(1-q)$ order, by \cite{tor-tir} \beq
\label{nFA} \left<n\right>_q = \left<n\right>_1 + (q-1)
\frac{(x-\psi)^2 e^{x-\psi}}{2 \left(e^{x-\psi}-1\right)^2}, \eeq
where $\left<n\right>_1$, $x$ and $\psi$ have the same definitions
as before. At this point, the remarkably simpler form of this
result, when compared to the result of the AA [Eq. (\ref{nAA})],
is worth emphasizing.\\


In this context, we have found general expressions for some
thermodynamical quantities of bosons and fermions \cite{tor-tir};
here we quote only the average number of particles for bosons,
since it will be adequate for our proposed comparison\footnote{We
take advantage here to signal out a mistake in the last equation
of Ref. \cite{tor-tir}, where the correction appears to be
proportional to 0.886 $(q-1)$ and should have a minus sign in
front of it \cite{WANG-PRIVA}.}: \beq \label{neFA}
\left<N_e\right>_q \le \frac {2\pi V(2mk)^{3/2} T^{3/2}} {h^3}
\left[2.315 + (q-1) 3.079 \right] . \eeq

\subsection{The exact result}

Although the results of the AA and  the FA have been successfully
used in a wide range of physical applications, an exact treatment
of non-extensive quantum statistics was lacking until the recent
work of Rajagopal, Mendes and Lenzi \cite{raj,ervin}. In their
analysis, they have given the many-particle $q$-Green function in
terms of a parametric contour integral over a kernel, multiplied
by the usual grand canonical one particle Green function which now
depends on $q$. They managed to obtain exact expressions for
thermodynamical quantities, such as $\left<N\right>_q$.\\

To proceed further, let us quote here  some of the results of
\cite{raj,ervin}. Rajagopal et al. have used the general contour
integral of the form, \beq b^{1-z} \frac{i}{2\pi} \int_C du
\exp(-bu) (-u)^{-z} = \frac{1}{\Gamma(z)} , \eeq with $b>0$ and Re
$z>0$, and where the contour $C$ starts from $+\infty$ on the real
axis, encircles the origin once counterclockwise and returns to
$+\infty$. Using the $q$-Green functions, and after some
cumbersome algebra, they finally obtain (for bosons) \beq
\left<N\right>_q = V \int_C du  K_q^{(2)}(u)
\int_{-\infty}^{\infty} \frac{d\omega}{2\pi} \int \frac{d^D
p}{(2\pi)^D} \frac{Z_1(-\beta(1-q)u,\mu)}
{\left[e^{-\beta(1-q)u(\omega-\mu)}-1\right]} A({\vec p};\omega),
\eeq where $D$ is the dimension of space, $ A({\vec p};\omega)$ is
the spectral weight function and \beq K_q^{(2)}(u) = i
\frac{\Gamma(1/(1-q))}{2\pi (Z_q)^q} \exp(-u)(-u)^{-1/(1-q)}, \eeq
and \beq Z_q(\beta,\mu) = \int_C du K_q^{(1)}(u) Z_1(-\beta
u(1-q),\mu). \eeq This exact expression for the average number of
particles finally gives us the opportunity to make a comparison
between the exact and the approximate results.

\section{Exact and approximate results}


\subsection{Bose-Einstein condensate}

One of possible experimental tests of the validity of the
$q$-framework is based on a recent work on Bose-Einstein
condensation of a small number of atoms (of the order of 100 to
170), confined to a small region of space by magnetic trapping
\cite{BE}. By taking free particle spectral weight function,
namely $A({\vec p};\omega)=2\pi\delta(\omega-{\vec p}^2/2m)$, near
the Bose-Einstein condensation, they have found

\bea \label{NqN1exact} \frac{\left<N\right>_q}{\left<N\right>_1}
\simeq \left[\frac{(T_c)_q}{(T_c)_1}\right]^{3/2}
\frac{\Gamma\left(\frac{2-q}{1-q}\right)} {(1-q)^{1/2}
\Gamma\left(\frac{2-q}{1-q} + \frac{1}{2}\right)} \left\{1 +
\frac{\left<N\right>_1}{(1-q)^{3/2}} \frac{\zeta(5/2)}{\zeta(3/2)}
\left[\frac{(T_c)_q}{(T_c)_1}\right]^{3/2} \right. \times
\nonumber\\ \times \left. \left[\frac{\Gamma\left(\frac{2-q}{1-q}
+ \frac{1}{2}\right)} {\Gamma\left(\frac{2-q}{1-q} + 2 \right)} -
q \frac{\Gamma\left(\frac{2-q}{1-q} \right)}
{\Gamma\left(\frac{2-q}{1-q} + \frac{3}{2}\right)} \right]
\right\}. \eea Here, the $\simeq$ sign reflects that this equation
is valid near the Bose-Einstein condensation, which does not
change the fact that it is an exact result. We may now expand this
expression in powers of $(q-1)$. Up to first order,

\beq \label{NqN1exactapprox}
\frac{\left<N\right>_q}{\left<N\right>_1} \simeq
\left[\frac{(T_c)_q}{(T_c)_1}\right]^{3/2} \left\{ 1 + (q-1)
\left( 0.456 - 0.023 \left<N\right>_1
\left[\frac{(T_c)_q}{(T_c)_1}\right]^{3/2} \right) \right\}. \eeq
The two previous equations deviate from each other very soon when
$(q-1)^2$ is not negligible.\\

Let us now derive similar expressions for the AA and FA in order
to compare them with Eqs. (\ref{NqN1exact}) and
(\ref{NqN1exactapprox}). The condition for the appearance of
Bose-Einstein condensation can be expressed as \beq
\left<N\right>_q > \left<N_e\right>_q . \eeq Alternatively, with
constant $\left<N\right>_q$ and $V$, using Eq. (\ref{neAA}) for
the AA and Eq. (\ref{neFA}) for the FA, this condition can be
recast in the form \beq T < \left(T_c\right)_q =
\frac{h^2}{(2\pi)^{3/2}2mk}
\left\{\frac{\left<N\right>_q}{V\left[2.315+(q-1)\kappa\right]}
\right\}^{2/3}, \eeq where $\kappa=4.27$ for the AA and
$\kappa=3.078$ for the FA. Then, we can organize these expressions
to give \beq \label{NqN1appr}
\frac{\left<N\right>_q}{\left<N\right>_1} =
\left[\frac{\left(T_c\right)_q}{\left(T_c\right)_1}\right]^{3/2}
\frac{[2.315+(q-1)\kappa]}{2.315} .
\eeq
Eq.(17) and (19) have
corrections which are trivial (not depending on $z$) just because
we have approximated them: the actual complete results are
Eqs.(15) and (16) for the AA, while those for the FA can be found
in our previous paper [24]. We managed the dependence on $z$ in
order to obtain an upper bound for the corrections and simplify
the analysis that follows. Differences between Eqs.
(\ref{NqN1appr}) and (\ref{NqN1exactapprox}) are worth noticing:
the later depends on $(T_c)_q$ and $\left<N\right>_1$ in a much
stronger way. However, as we shall see, for $q$ close to 1 these
differences are not important.\\

We would now like to choose physically suitable $q$ values. An
early Universe test based on the FA has shown \cite{early-FA} to
produce a bound $|q-1|\le 4.01 \times 10^{-3}$, thus we have
$q=0.996$. The other $q$ value which we use comes from a very
recent work on pion transverse-momentum correlations in Pb-Pb
high-energy nuclear collisions \cite{pion}. In that work, a
deviation of $|q-1|=0.015$ from the standard statistics is found
to be sufficient for eliminating the puzzling discrepancy between
theoretical calculations and experimental data \cite{pion}. Thus,
we shall use $q=0.985$ (in fact, in \cite{pion}, $q=1.015$ has
been used, but since the exact result is given for $q<1$ values,
we must take $q=0.985$, which has the same $|q-1|$ deviation). In
Fig. 3 we plot $\left<N\right>_q / \left<N\right>_1$ versus
$\left(T_c\right)_q / \left(T_c\right)_1$ for two representative
values of $\left<N\right>_1$, and the two quoted values of  $q$.
However, note again that in our approximated schemes,
$\left<N\right>_q / \left<N\right>_1$ as a function of
$\left(T_c\right)_q / \left(T_c\right)_1$ is in fact independent
of the particular value of $ \left<N\right>_1$.
From Fig. 3, the following conclusions can be drawn: At the order
of such $q$ values, the AA and the FA are almost the same, and in
$(1-q)$-order correction, any of them could be used with the same
confidence (maybe the FA would be preferable due to its remarkably
simpler form). Only in those situations of extremely high
experimental precision one could distinguish between the exact and
approximate results.

\subsection{Blackbody radiation}

Very recently, an exact analysis of the blackbody radiation within
the $q$-framework has been given \cite{bb-exact}. This exact
analysis gives the generalization of the Stefan-Boltzmann law as
\beq \label{Uexact} U_q =\frac{3kT\xi_3}{Z_q^q}
\sum_{m=0}^{\infty} \frac{\xi_3^m}{m!}
\frac{\Gamma[(2-q)/(1-q)]}{\Gamma[(2-q)/(1-q)+3(m+1)]}, \eeq where
\beq Z_q = \sum_{m=0}^{\infty} \frac{\xi_3^m}{m!}
\frac{\Gamma[(2-q)/(1-q)]}{\Gamma[(2-q)/(1-q)+3m]} , \eeq and \beq
\xi_3 = \frac{4 \Gamma(3)
\zeta(4)}{[2\pi^{1/2}(1-q)]^3\Gamma(3/2)} \left(\frac{2\pi
V^{1/3}kT}{hc}\right)^3 . \eeq Let us now recall the
Stefan-Boltzmann law derived by using the AA
\cite{SB-AAplastino,SB-AAwang} and the FA \cite{SB-FAugur}: \beq
\label{Uappr} U_q = \frac{8 \pi k^4 T^4 V}{c^3 h^3} \left[6.4939 -
(1-q)\; \theta\; \right] \eeq where $\theta = 40.018$ for the AA
and $\theta = 62.215$ for the FA.\\

For the comparison of the exact and approximate Stefan-Boltzmann
laws, we again try to choose a value of $q$ which is in accordance
with the blackbody radiation. The possible $q$-correction could be
at the order of $10^{-4}$ or $10^{-5}$. Thus, here we shall use
again the largest deviation predicted for the $q$-correction
\cite{SB-AAplastino}, namely $|q-1|\le 5.3 \times 10^{-4}$, which
gives $q=0.99947\;$. In Fig. 4 we present the behaviour of the
exact [Eq. (\ref{Uexact})] and the approximate results [Eq.
(\ref{Uappr})] for $q=0.99947$. It is seen from the figure that
for such order of $q$-correction the approximate results are very
close to the standard ($q=1$) case without exhibiting any
curvature, contrary to the exact result.

\section{Final remarks}

We have managed to develop an analytical technique to express
thermodynamical quantities for the asymptotic approach of quantum
distribution functions. We have shown that, for simple boson
systems, and for all $q$-values admitted by the existing bounds,
both approximate schemes (the AA and the FA) are in agreement with
the exact result (see figures). The magnitude of the deviation is
quantified in previous formulae and could be seen if there is
enough experimental precision. Otherwise, the simpler form that
the factorization approach exhibits makes a case for its use as a
standard and safe procedure for $(1-q)$-order corrections.

\section*{Acknowledgments}
U.T. acknowledges the partial support of BAYG-C program of TUBITAK
(Turkish Agency) and CNPq (Brazilian agency) as well as the
support from Ege University Research Fund under the project number
98FEN025. D.F.T. acknowledges support from CONICET and Fundaci\'on
Antorchas and wishes to thank A. Lavagno for sending him his
valuable Ph. D. Thesis. We also thank A. K. Rajagopal, A. Wang, A.
Erzan and an anonymous referee for useful comments.


\newpage

\section*{Figure Captions}

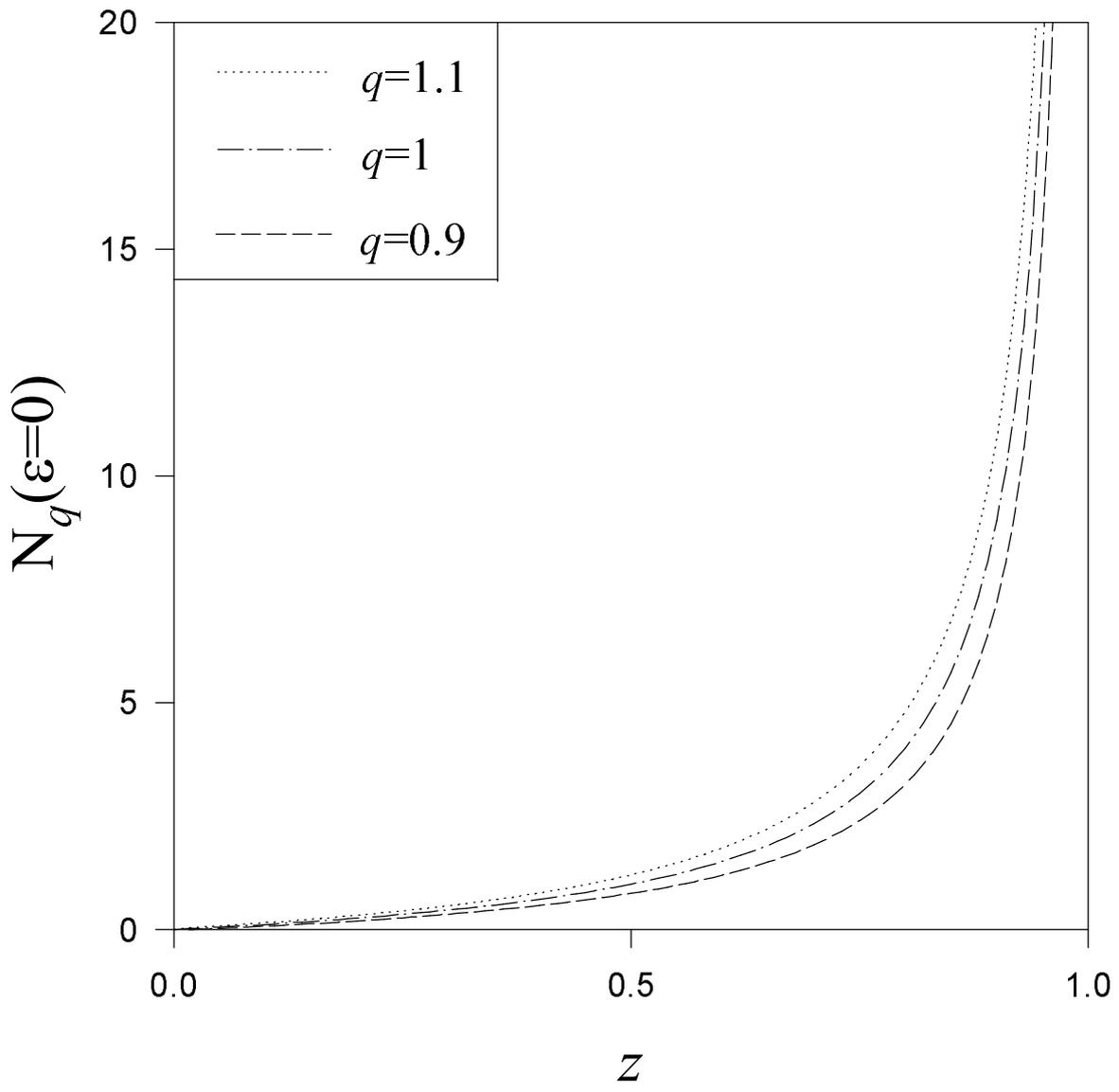
Figure 1 : $N_q(\epsilon=0)$ of the AA as a function of $z$
for different $q$ values.\\

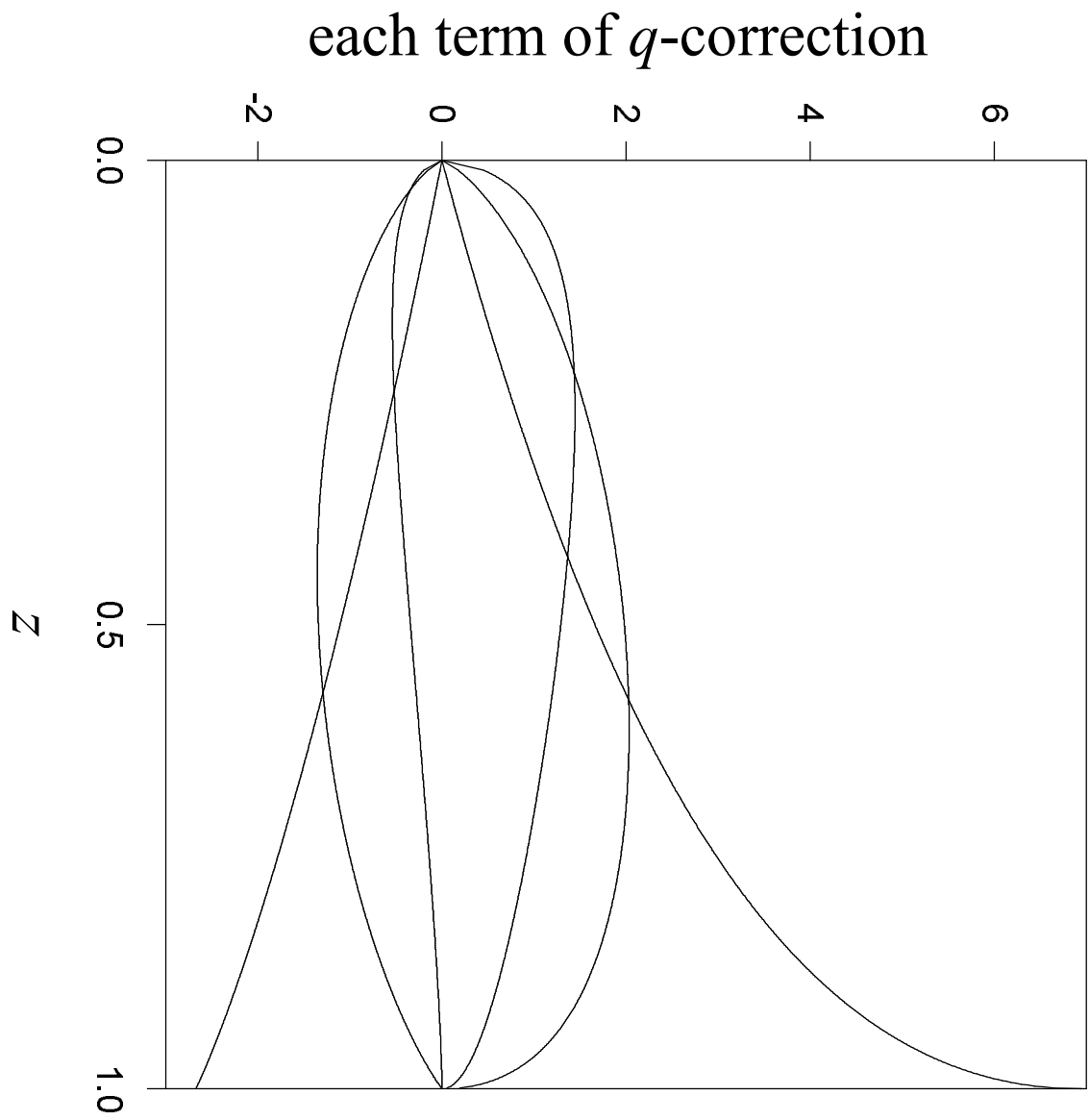
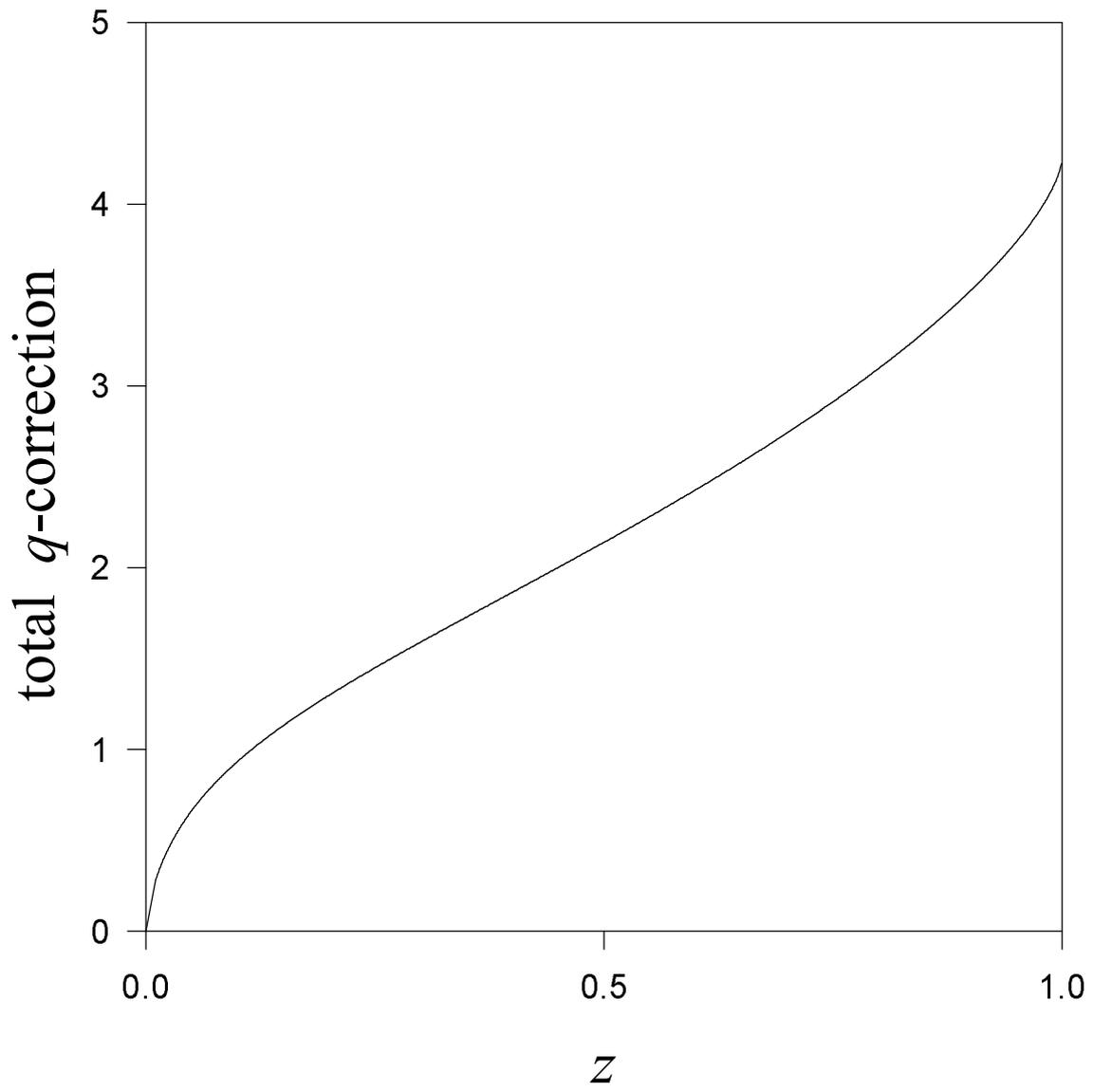
Figure 2 : {\bf (a)} The contribution of the
different terms that enter the $(q-1)$ correction to the average
number  of bosons within the AA.
On the right corner of the figure, the curves
corresponds to the following order:
first, fourth, sixth, fifth, second and third term.
{\bf (b)} The total $(q-1)$ correction to the average number of
 bosons within the AA. Its maximum possible value is
attained at $z=1$, and the correction goes as
$\pi^{1/2} (3/2 \zeta(3/2)- 9/8 \zeta(5/2))=4.27\;$.\\

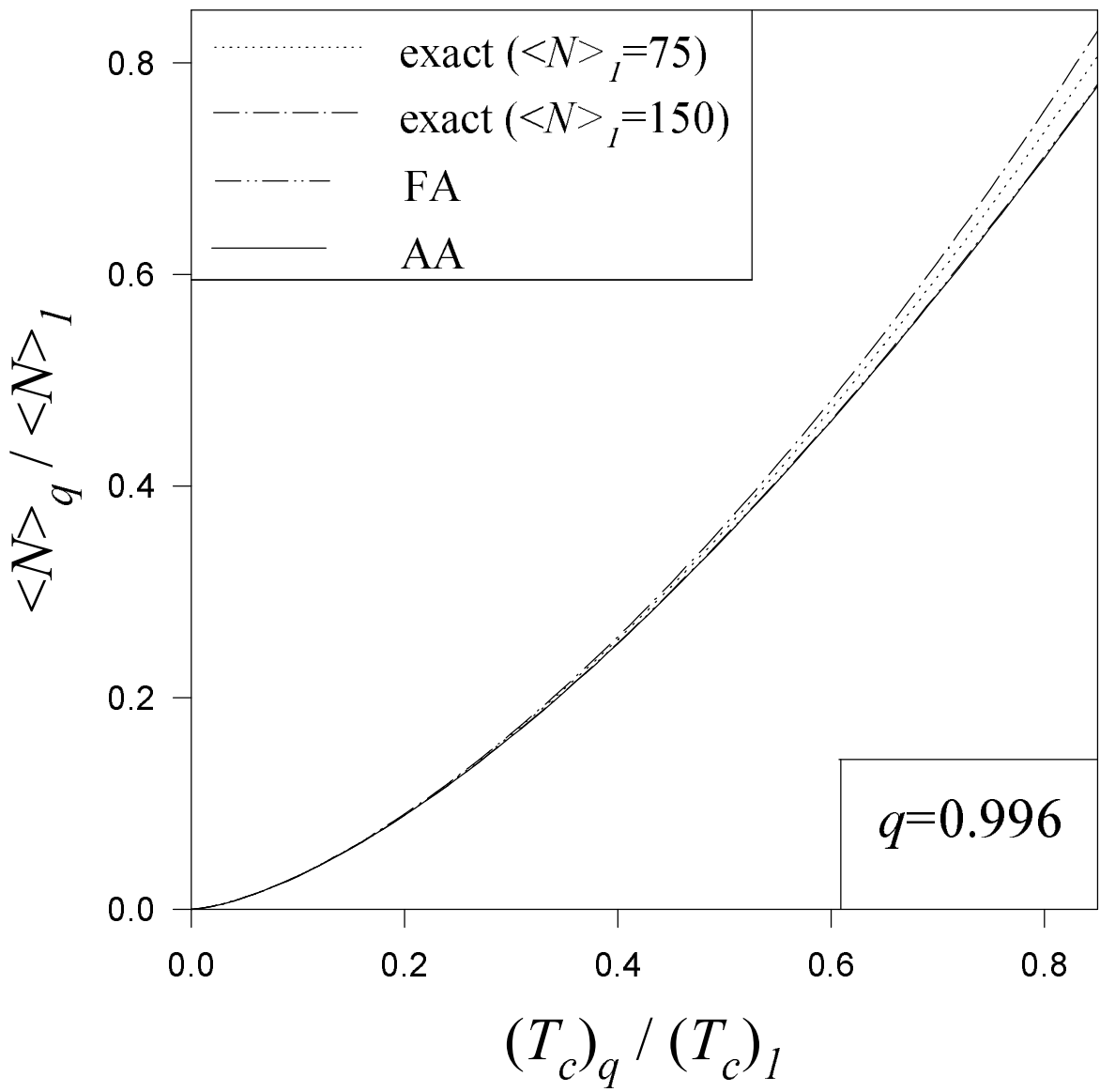
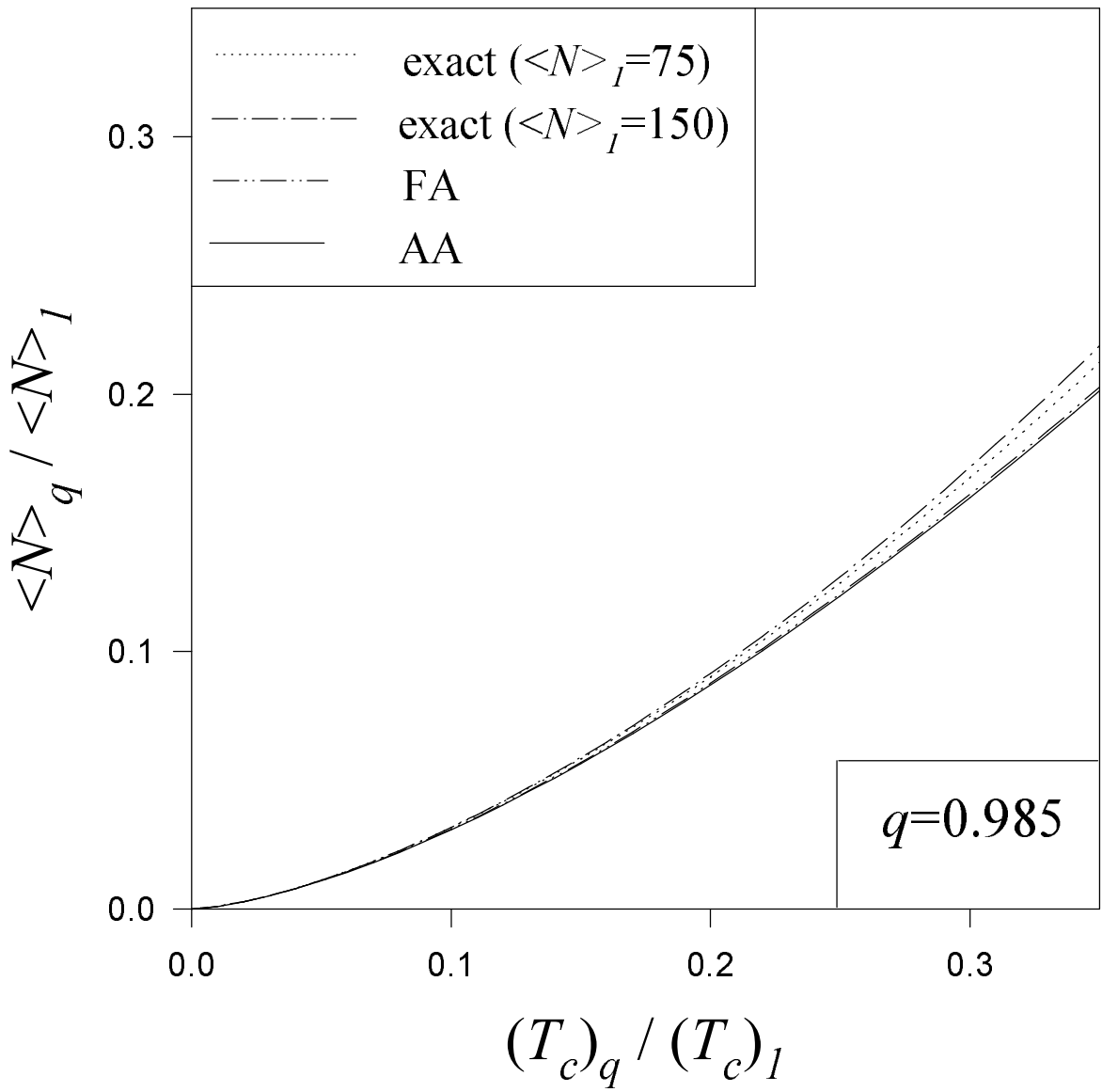
Figure 3 : Bose-Einstein condensation: Plot of
$\left<N\right>_q / \left<N\right>_1$ as a function of
$\left(T_c\right)_q / \left(T_c\right)_1$
for {\bf (a)} $q=0.996$ and {\bf (b)} $q=0.985\;$.\\

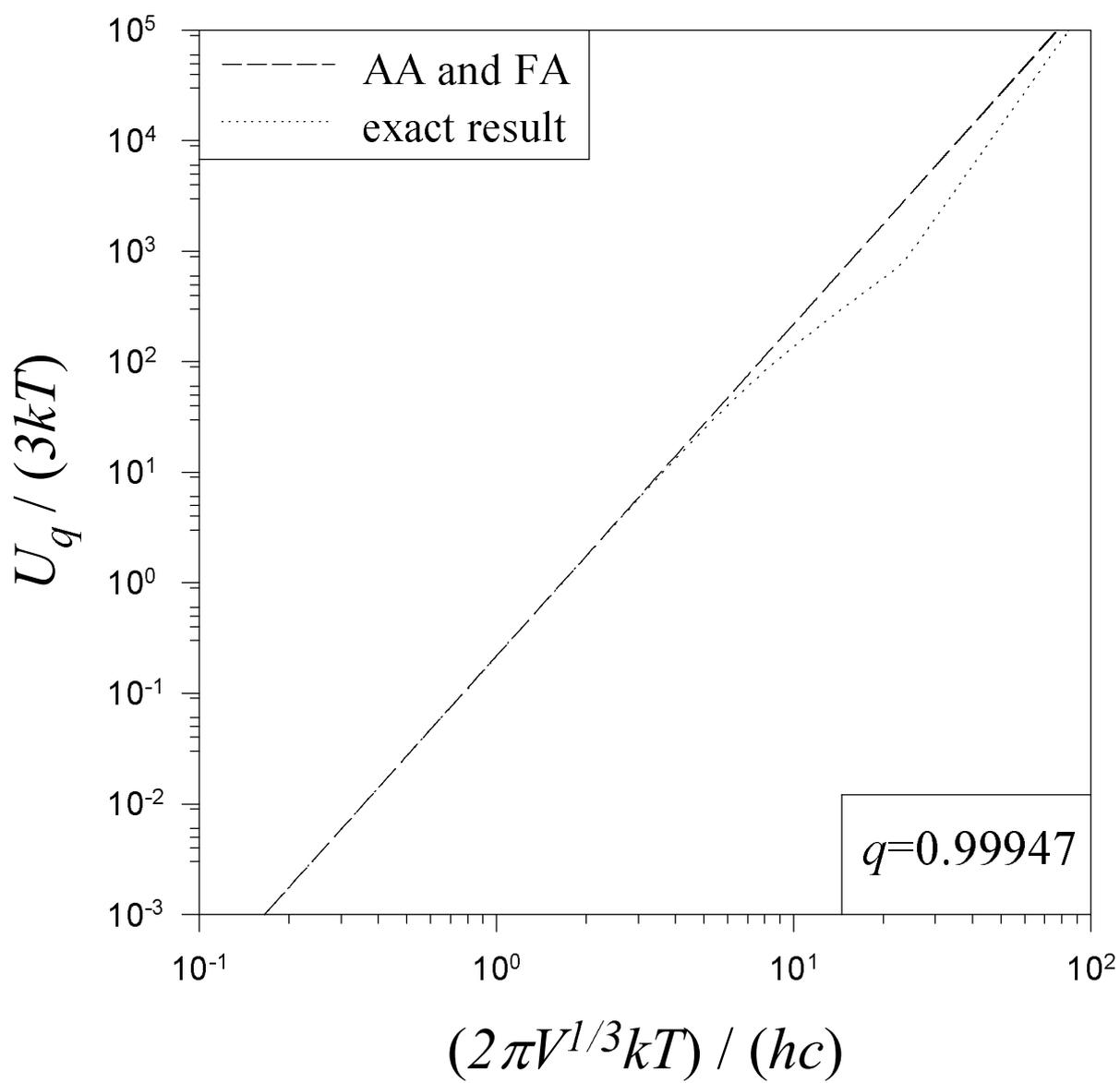
Figure 4 : Blackbody radiation: Internal energy versus
$(2\pi kT)/(hc)$ for $q=0.99947\;$.\\

\newpage













\appendix
\section{}
The calculation of the integral $a$ can be done as follows. Let us
define, introducing an extra parameter $m$, \beq I =
\int_0^{\infty} \frac{(x-\psi) x^{1/2} dx} {e^{x-\psi}-m}. \eeq
Then, we could write, \beq a = \left[ \frac{dI}{dm}\right]_{m=1} =
\left[ \int_0^{\infty} \frac{(x-\psi) x^{1/2} dx}
{\left(e^{x-\psi}-m\right)^2}\right]_{m=1}. \eeq Defining
$m^{-1}\equiv e^{\phi}$, it is easy to write, \beq a_m =
\frac{d}{dm} \int_0^{\infty} \frac{(x-\psi) x^{1/2} dx} {m
\left(m^{-1} e^{x-\psi}-1\right)}= \frac{d}{dm}\left\{\frac{1}{m}
\left[\int_0^{\infty} \frac{x^{3/2} dx} {e^{x-\psi^{\prime}}-1} -
\psi \int_0^{\infty} \frac{x^{1/2} dx}
{e^{x-\psi^{\prime}}-1}\right]\right\}, \eeq where
$\psi^{\prime}\equiv \psi - \phi$. This let us to obtain, \beq a_m
= \frac{d}{dm}\left[\frac{1}{m} \left( \Gamma(5/2)
g_{5/2}(\psi^\prime) - \psi \Gamma(3/2)
g_{3/2}(\psi^\prime)\right) \right]. \eeq
Finally this gives us
the solution of the integral $a$: \beq a = \Gamma(5/2) g_{3/2}(z)
- \Gamma(5/2) g_{5/2}(z) - \psi \Gamma(3/2) g_{1/2}(z) + \psi
\Gamma(3/2) g_{3/2}(z) . \eeq For the calculation of $c$, we may
conveniently define $a_l$ as, \beq a_l=\int \frac{ (x-\psi)
x^{1/2} dx } {\left[e^{l(x-\psi)} -1\right]^2 } \eeq and derive
with respect to $l$ to obtain \beq \left[ \frac{da_l}{dl}
\right]_{l=1} = -2 \int \frac{ (x-\psi)^2 x^{1/2} e^{x-\psi} dx }
{ (e^{x-\psi} -1 )^3 }=-2 c . \eeq Now, to compute $a_l$, we may
change variables as follows: \beq \tilde x=lx, \hspace{2.5cm}
\tilde \psi=l\psi . \eeq We then obtain, \beq a_l=\frac
{1}{l^{5/2}} \int \frac{ (\tilde x-\tilde \psi) \tilde x^{1/2}
d\tilde x } { (e^{(\tilde x-\tilde \psi)} -1 )^2 }= \frac
{1}{l^{5/2}}\left\{ \Gamma(5/2) \left[ g_{3/2}(\tilde \psi) -
g_{5/2}(\tilde \psi) \right] + \tilde \psi \Gamma(3/2) \left[
g_{3/2}(\tilde \psi) -g_{1/2}(\tilde \psi) \right] \right\} . \eeq
Deriving with respect to $l$, we get \bea \frac{da_l}{dl} = -
\frac {5}{2l^{7/2}} \left\{ \Gamma(5/2) \left[ g_{3/2}(\tilde
\psi) - g_{5/2}(\tilde \psi) \right] + \tilde \psi \Gamma(3/2)
\left[ g_{3/2}(\tilde \psi) -g_{1/2}(\tilde \psi) \right] \right\}
+             \nonumber \\ \frac {1}{l^{5/2}}\left\{ \Gamma(5/2)
\left[ \psi g_{1/2}(\tilde \psi) - \psi g_{3/2}(\tilde \psi)
\right] + \psi \Gamma(3/2) \left[ g_{3/2}(\tilde \psi)
-g_{1/2}(\tilde \psi) \right]  \right.\nonumber \\ +  \left.
\tilde \psi \Gamma(3/2) \left[ \psi g_{1/2}(\tilde \psi) -\psi
g_{-1/2}(\tilde \psi) \right] \right\} \eea \mbox{} From this
equation, the integral $c$ can be obtained by making $l=1$ and
$\tilde \psi=\psi$. Here, we should note that, in all previous
calculations, we have used (i) the result \beq
g_{n-1}(z)=z\frac{\partial}{\partial z}\left[g_n(z)\right]=
\frac{\partial}{\partial \ln (z)}\left[g_n(z)\right]. \eeq and
(ii) Robinson's power series representation \cite{ROBIN} (which is
valid for {\it all $n$ values}): \beq g_n(\alpha)=\Gamma(1-n)
\alpha^{n-1} + \sum_{l=0}^\infty \frac{(-)^l}{l!}
\zeta(n-l)\alpha^l, \eeq for the $g_n$ functions, where
$\alpha=-\ln(z)$ and $\zeta$ is the Riemann zeta function. Using
this result, one can recover the relationship for the derivatives
of $g_n$ functions, and use it to evaluate $g_{-1/2}$. We finally
obtain the result for the integral $c$: \bea c = \frac{5}{4}
\Gamma(5/2) g_{3/2}(\psi) - \frac{5}{4} \Gamma(5/2) g_{5/2}(\psi)
+ \frac{5}{4} \Gamma(3/2) \psi g_{3/2}(\psi) - \frac{5}{4}
\Gamma(3/2) \psi g_{1/2}(\psi) \nonumber \\ - \frac{1}{2}
\Gamma(5/2) \psi g_{1/2}(\psi) + \frac{1}{2} \Gamma(5/2) \psi
g_{3/2}(\psi) - \frac{1}{2} \Gamma(3/2) \psi g_{3/2}(\psi) +
\frac{1}{2} \Gamma(3/2) \psi g_{1/2}(\psi) \nonumber \\ -
\frac{1}{2} \Gamma(3/2) \psi^2 g_{1/2}(\psi) + \frac{1}{2}
\Gamma(3/2) \psi^2 g_{-1/2}(\psi). \eea We proceed further to
compute integral $b$. To do so, we apply the following procedure.
Let us define a new integral with an extra parameter, such that
\beq a_m = \int_0^{\infty} \frac{ x^{1/2} dx}
{e^{m(x-\psi)}-1}=\frac{1}{m^{3/2}} \int_0^{\infty} \frac{ \tilde
x^{1/2} d\tilde x} {e^{(\tilde x-\tilde
\psi)}-1}=\frac{1}{m^{3/2}} \Gamma(3/2) g_{3/2}(\tilde \psi) ,
\eeq where we have used the change of variable $\tilde x = m x$
and $\tilde \psi = m \psi$. If we now derive with respect to $m$,
we obtain, \beq \left[ \frac{da_m}{dm} \right]_{m=1} = \left[ -
\int_0^{\infty} \frac{ (x-\psi) x^{1/2} e^{m(x-\psi)} dx}
{(e^{m(x-\psi)}-1)^2} \right]_{m=1}=-b. \eeq Since we have, \beq
\frac{da_m}{dm} = - \frac{3}{2m^{5/2}} \Gamma(3/2) g_{3/2}(\tilde
\psi) + \frac{1}{m^{3/2}}  \Gamma(3/2) \psi g_{1/2}(\tilde \psi) ,
\eeq it is easy to write down the solution of the integral $b$:
\beq b= \frac{3}{2} \Gamma(3/2) g_{3/2}(\psi) - \Gamma(3/2) \psi
g_{1/2}(\psi) . \eeq Once the integral $b$ is calculated, the
integral $d$ can be obtained as follows. Let us define $a_m$ as
above. Deriving it twice with respect to the parameter $m$, we
obtain \beq \left[ \frac{d^2a_m}{dm^2} \right]= 2d - I_{new} ,
\eeq where, \beq I_{new}= \int_0^{\infty} \frac{(x-\psi)^2 x^{1/2}
e^{m(x-\psi)} dx} {\left[e^{m(x-\psi)}-1\right]^2} . \eeq It is
easy to compute this integral with a similar trick. We need to
define, with usual notation, \beq I_{ext}= \int_0^{\infty}
\frac{(x-\psi) x^{1/2}  dx} {\left[e^{l(x-\psi)}-1\right]} =
\frac{1}{l^{5/2}} \left[ \Gamma(5/2) g_{5/2}(\tilde\psi) -\tilde
\psi \Gamma(3/2) g_{3/2}(\tilde\psi)\right] \eeq and derive it
with respect to $l$. Further evaluation in $l=1$ reproduces
$I_{new}$: \beq I_{new}= \frac{5}{2} \Gamma(5/2) g_{5/2}(\psi) -
\psi \frac{3}{2} \Gamma(3/2) g_{3/2}(\psi) - \psi  \Gamma(5/2)
g_{3/2}(\psi) + \psi^2 \Gamma(3/2) g_{1/2}(\psi). \eeq Thus, we
finally have the solution of the integral $d$: \bea d =
\frac{15}{8} \Gamma(3/2) g_{3/2}(\psi) - \frac{3}{2} \psi
\Gamma(3/2) g_{1/2}(\psi) + \frac{3}{4} \psi^2 \Gamma(3/2)
g_{-1/2}(\psi) +\frac{5}{4} \Gamma(5/2) g_{5/2}(\psi) - \nonumber
\\ \psi \frac{3}{4} \Gamma(3/2) g_{3/2}(\psi) - \frac{1}{2}\psi
\Gamma(5/2) g_{3/2}(\psi) + \frac{1}{2}\psi \Gamma(3/2)
g_{3/2}(\psi) + \frac{1}{2}\psi^2 \Gamma(3/2) g_{1/2}(\psi) . \eea

\end{document}